\documentclass[pre,twocolumn,showpacs,superscriptaddress,preprintnumbers,floatfix]{revtex4}

\usepackage{dcolumn}
\usepackage{url}
\usepackage{amsmath}
\usepackage{amssymb}
\usepackage{graphicx}
\usepackage{bm}   
\usepackage{bbm}
\usepackage{verbatim}
\usepackage{stmaryrd}
\usepackage{amsthm}
\newcommand\degrees[1]{\ensuremath{#1^\circ}}

\begin{document}

\title{Insects, Trees, and Climate:\\
The Bioacoustic Ecology of Deforestation and\\
Entomogenic Climate Change}

\author{David Dunn}
\email{artscilab@comcast.net}
\affiliation{Art and Science Laboratory, Santa Fe, New Mexico 87501}

\author{James P. Crutchfield}
\email{chaos@cse.ucdavis.edu}
\affiliation{Center for Computational Science \& Engineering and Physics Department,
University of California Davis, One Shields Avenue, Davis, CA 95616}
\affiliation{Art and Science Laboratory, Santa Fe, New Mexico 87501}

\date{\today}

\bibliographystyle{unsrt}

\begin{abstract}
Accumulating observational evidence suggests an intimate connection between
rapidly expanding insect populations, deforestation, and global climate change.
We review the evidence, emphasizing the vulnerability of key planetary carbon
pools, especially the Earth's forests that link the micro-ecology of insect
infestation to climate. We survey current research regimes and insect control
strategies, concluding that at present they are insufficient to cope with the
problem's present regional scale and its likely future global scale. We propose
novel bioacoustic interactions between insects and trees as key drivers of
infestation population dynamics and the resulting wide-scale deforestation. The
bioacoustic mechanisms suggest new, nontoxic control interventions and
detection strategies.
\end{abstract}


\preprint{Santa Fe Institute Working Paper 06-12-XXX}
\preprint{arxiv.org/q-bio.PE/0612XXX}

\maketitle

\tableofcontents

\section{Introduction}

Forest ecosystems result from a dynamic balance of soil, insects, plants, animals, and
climate. The balance, though, can be destabilized by outbreaks of tree-eating insects.
These outbreaks in turn are sensitive to climate, which controls precipitation. Drought
stresses trees, rendering them vulnerable to insect predation. The net result is increased
deforestation driven by insects and modulated by climate.

For their part, many predating insects persist only to the extent they successfully
reproduce, which they do by consuming and living within trees. Drought-stressed trees
are easier to infest compared to healthy trees, which have more robust defenses
against attack. To find trees suitable for reproduction, insects track relevant
environmental indicators, including chemical signals and, possibly, bioacoustic
ones emitted by stressed trees. At the level of insect populations, infestation dynamics
are sensitive to climate via seasonal temperatures. Specifically, insect populations
increase markedly each year when winters are short and freezes less severe. The net
result is rapidly changing insect populations whose dynamics are modulated by climate.

Thus, via temperature and precipitation, climate sets the context for tree growth and
insect reproduction and also for the interaction between trees and insects. At the largest
scale, climate is driven by absorbed solar energy and controlled by relative fractions
of atmospheric gases. The amount of absorbed solar energy is determined by cloud and ground
cover. Forests are a prime example, as an important ground cover that absorbs, uses, and
re-radiates solar energy in various forms. At the same time forests are key moderators of
atmospheric gases. Trees exhaust oxygen and take up carbon dioxide in a process that
sequesters in solid form carbon from the atmosphere. As plants and trees evolved, in fact,
they altered the atmosphere sufficiently that earth's climate, once inhospitable, changed
and now supports a wide diversity of life.

There are three stories here: the trees', the insects', and the climate's. They necessarily
overlap since the phenomena and interactions they describe co-occur in space and in
time. Their overlap hints at an astoundingly complicated system, consisting of many
cooperating and competing components; the health of any one depending on the health of
others. (Figure 1 gives a schematic view of these components and their interactions.)
How are we to understand the individual views as part of a larger whole? In particular,
what can result from interactions between the different scales over which insects, trees,
and climate adapt?

Taking the stories together, we have, in engineering parlance, a \emph{feedback loop}: Going
from small to large scale, one sees that insects reproduce by feeding on trees, forests
affect regional solar energy uptake and atmospheric gas balance, and, finally, energy
storage and atmospheric gases affect climate. Simultaneously, the large scale (climate)
sets the context for dynamics on the small scale: temperature modulates insect
reproduction and precipitation controls tree growth. The feedback loop of insects, trees,
and climate means that new kinds of behavior can appear---dynamics not due to any
single player, but to their interactions. Importantly, such feedback loops can maintain
ecosystem stability or lead to instability that amplifies even small effects to large scale.

\begin{figure}
\begin{center}
\resizebox{!}{2.50in}{\includegraphics{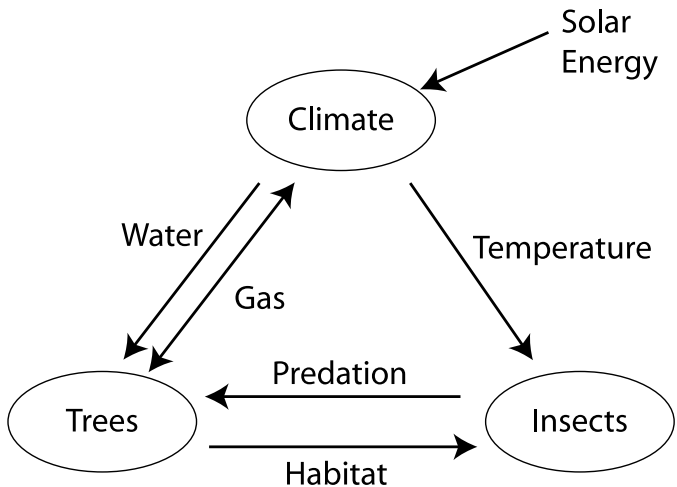}}
\end{center}
\caption{Interaction of insects, trees, and climate.}
\label{fig:Interactions}
\end{figure}

Here we give a concrete example of the dynamic interaction between insects, trees, and climate.
We focus on the role that bark beetles (Scolytidae or, more recently, Curculionidae: Scolytinae)
play in large-scale deforestation and consequently in climate change. Bark beetles are emblematic
of many different insect species that now participate in rapid deforestation. Likewise, we primarily
focus upon the North American boreal forests for their unique characteristics but also as
representative of the vulnerability of all types of forest ecosystems. And so, the picture we
paint here is necessarily incomplete. Nonetheless, their cases serve to illustrate the complex
of interactions that are implicated in the feedback loop and also the current limits to human
response. Although they are not alone, bark beetles appear to be an example of a novel
player in climate change. Unlike the climatic role that inanimate greenhouse gases are
predicted to play in increasing global temperature over the next century, bark beetles
represent a biotic agent that actively adapts on the time scale of years but that, despite the
short time scale, still can cause effects, such as deforestation, at large spatial scales. To
emphasize the novelty of this kind of biological, non-human agent, we refer to the result
as \emph{entomogenic climate change}.

In analyzing the relationship between the feedback loop components, one important
conclusion is that we understand relatively little about the interactions between insects
and trees and the dynamics of infestation. In particular, we see that there is a need to
expand on the success of, and to acknowledge the limitations of, the dominant chemical
ecology model of insect infestation. 

A detailed analysis of the problem of entomogenic climate change leads us to make a
number of constructive suggestions for increased attention to relatively less familiar
domains of study, including micro-ecological symbiosis and its nonlinear population
dynamics, and insect social organization. Here we emphasize in particular the role that
bark beetle bioacoustic behavior must have in their evolving multiple survival
adaptations which, it appears, fill in significant gaps in the explanatory model of
infestation dynamics. One goal is to stimulate interdisciplinary research that is
appropriate to the complex of interactions implicated in deforestation and appropriate
to discovering effective control strategies.

\section{Forest Health and Climate: A recent snapshot}

The Earth's three great forest ecosystems---tropical, temperate, and boreal---are of
irreplaceable importance to its self-regulating balance. Their trees help to regulate its
climate, provide essential timber resources, and create a diversity of habitat and nutrients
that support other forms of life, including millions of people. Forests contribute to global
climate dynamics through a carbon cycle in which atmospheric carbon dioxide is converted
into an immense carbon pool. At any one point in time, the Earth's forest ecosystems together 
hold a majority of the Earth's carbon stocks with the boreal forests comprising 49 percent of 
the total carbon pool contained within these three types of forest ecosystems \cite{Malh99a}. 
That carbon is then slowly released back into the atmosphere through complicated decomposition 
processes. 

All forms of deforestation, human and natural, directly impact climatic conditions by
attenuating or delaying the carbon cycle. In concert with well-documented greenhouse
gas effects that drive global atmospheric change, the potential loss of large areas of these forests, combined with accelerating deforestation of tropical and temperate regions, may have significant future climate impacts beyond already dire predictions. Ice core studies have revealed
that the Earth's climate has varied cyclically over the past 450,000 years. Temperatures have
been closely tied to variations in atmospheric carbon dioxide in a cyclic change that recurs
on the time scale of millennia. Vegetation has been forced to adapt. The boreal forests are,
in fact, highly vulnerable to these climate shifts. Examination of fossil pollen and other
fossil records shows that, in response to temperature variations over the past millennia,
North American boreal forests have radically changed many times \cite{Lind02a}. The unique
sensitivity of these forests' tree species to temperature suggests that the predicted warmer
climate will cause their ecological niches to shift north faster than the forests can migrate.
Researchers believe that, in addition to other deforestation factors, the boreal forests may
eventually be substantially reduced to just half their current size over the next
century \cite{Smit00a}.

One major consequence of boreal deforestation is increasing fire risk. Even though
forests require fire for reproduction and rejuvenation, a warmer climate will most likely
push an otherwise natural disturbance to an extreme frequency and scale. Over the next
half-century, the Siberian and Canadian boreal forests will most likely see as much as a
50 percent increase in burnt trees \cite{Smit00a}. One of the major sources fueling these fires will be
dead and dying trees killed by various opportunistic insect species and their associated
microorganisms.

Paralleling the concerns about the boreal forests, in recent years there has been a growing
awareness of extensive insect outbreaks in various regional forests throughout the
western United States. At first, local and national media reported on these outbreaks, and
the surprising devastation to forested areas, as the result of regional drought conditions
that encouraged various species of bark beetle to thrive. As consecutive summers of
unprecedented forest fires consumed the dead and dying trees, though, a new concern
emerged: insect-driven deforestation is a much larger threat connected to global climate
change. In fact, climate experts, forestry personnel, and biologists, have all observed that
these outbreaks are an inevitable consequence of a climatic shift to warmer temperatures \cite{Smit00a}.

Biologists are now voicing concern that the problem exceeds any of the earlier
projections. Evidence from diverse research sources suggests we are entering an
unprecedented planetary event: forest ecology is rapidly changing due to exploding
plant-consuming (\emph{phytophagous}) insect populations. In 2004, NASA's Global Disturbances
project analyzed nineteen years of satellite data ending in 2000. It revealed rapid
defoliation over a brief period (1995 to 2000) of a vast region that extends from the
US-Canadian border in western Canada to Alaska. The conclusion was that the
devastation resulted from two different insects, the mountain pine beetle (\emph{Dendroctonus
ponderosae}) and the western spruce budworm (\emph{Choristoneura occidentalis})
\cite{Cohe04a}. Ecologist
Chris Potter (NASA Ames Research Center) said at the time \cite{Cohe04a}: ``This looks like
something new happening on a huge scale. It's a sudden shift into a new kind of forest
condition.''

Now, two years later we know of even further damage. In Alaska, spruce bark beetles
(\emph{Dendroctonus rufipennis}) have killed 4.4 million acres of forest
\cite{Joli06a}. This damage results
from only one such insect. Alaska is also witnessing population explosions of many
others, including the western spruce budworm, the black-headed budworm (\emph{Acleris
gloverana Walsingham}), the amber-marked birch leaf miner (\emph{Profenusa thomsoni}), and
the aspen leaf miner (\emph{Phyllocnistis populiella}). In British Columbia the mountain pine
beetle has infested 21 million acres and killed 411 million cubic feet of trees. This is
twice the annual take by all the loggers in Canada. The general consensus is that beetles
will soon take 80 percent of the pines in the central forest of British Columbia.  The Canadian
Forest Service now calls the beetle invasion of Canada the largest known insect
infestation in North American history \cite{Stru06a}.

Jesse Logan (USDA Forest Service) and James Powell (Utah State University, Logan)
discussed the serious implications that a continuing warming trend will have on the
range expansion of the mountain pine beetle into both higher elevations and more
northern latitudes \cite{Loga01a}. At the time, one concern was that the beetles would breach the
Canadian Rockies and expand into the great boreal forests of Canada. Historically, these
forests have been immune to beetles due to predictably severe winter conditions that
greatly attenuate beetle populations. Since much of Canada has seen mean winter
temperature increases as high as $\degrees{4}$C in the last century, and even faster changes recently,
the conditions for the beetles are improving rapidly. As British Columbia forestry officer
Michael Pelchat recently said \cite{Stru06a}: ``We are seeing this pine beetle do things that have
never been recorded before. They are attacking younger trees, and attacking timber in
altitudes they have never been before.''

It is now well established that mountain pine beetles have slipped through mountain
passes from the Peace River country in northern British Columbia to Alberta, the most
direct corridor to the boreal forests. If the beetle is successful at adapting to and
colonizing Canada's jack pine, there will be little to stop it moving through the immense
contiguous boreal forest, all the way to Labrador and the North American east coast. It
then will have a path down into the forests of eastern Texas. Entomologist Jesse Logan
\cite{Loga01a} describes this as ``a potential geographic event of continental scale with unknown, but
potentially devastating, ecological consequences.''

Continental migration aside, if the beetles infest the high-elevation conifers, the so-called
five-needle pines, of the western United States, this will reduce the snow-fence effect that
these alpine forests provide. Snow fences hold windrows of captured snow that are
crucial to the conservation and distribution of water from the Rocky Mountains. This is
one of the primary origins of water that sources several major river systems in North
America \cite{Loga01a}. The Rocky Mountains and the Southwest have seen massive die off of
ponderosa and pinion pines numbering in the millions due to bark beetle infestation.
Every western state is contending with various rates of unprecedented insect infestation
not only by many different species of Scolytidae, but also by other plant-eating insects.

While the conifers of the boreal forests have been the most dramatically affected, many temperate forest tree species are also
struggling. The emerald ash borer (\emph{Agrilus planipennis}) has killed over 20 million ash
trees in Michigan, Ohio, and Indiana. In 2006 it was observed to have moved into
northern Illinois and Ontario, Canada \cite{Carl06a}. A large, wood-boring conifer pest, the sirex
woodwasp (\emph{Sirex noctilio})---native to Europe and Asia---has now entered several New
York counties and southern Canada. It has recently devastated millions of pine trees in
Australia, South America, and South Africa \cite{Carl06a}. These and other rising populations of
phytophagous insects are now becoming recognized as a global problem and one of the
most obvious and fast emerging consequences of global climate change. Over the past
fifteen years there have been reports of unusual and unprecedented outbreaks occurring
on nearly every continent.

\section{What Drives Infestations?}

Several well-understood factors underlie how climate change impacts insect populations.
The two dominant environmental factors are changes in temperature and moisture.
Changing insect-host relationships and nonhost species impacts, such as predation and
disease, also play essential roles.

Since insects are cold-blooded (\emph{poikilothermic}), they are extremely sensitive to
temperature, being more active at higher temperatures. As winter temperatures increase,
there are fewer freezing conditions that keep insect populations in check than in the past.
Shortened winters, increasing summer temperatures, and fewer late-spring frosts correlate
to increased insect feeding, faster growth rates, and rapid reproduction.

Moisture availability and variability are also major determinants of insect habitat---forest
health and boundaries. Drought creates many conditions that are favorable to increased
insect reproduction. Many drought-induced plant characteristics are attractive to insects.
Higher plant surface temperatures, leaf yellowing, increased infrared reflectance, biochemical
 changes, and possibly stress-induced cavitation acoustic emissions, may all be
positive signals to insects of host vulnerability. Drought also leads to increased food
value in plant tissues through nutrient concentration, while reducing defensive
compounds. These last factors may in turn increase the efficacy of insect immune
systems and therefore enhance their ability to detoxify remaining plant defenses. Higher
temperatures and decreased moisture may also decrease the activity of insect diseases and
predator activity while optimizing conditions for mutualistic microorganisms that benefit
insect growth \cite{Matt87a}.

One of the most frequently noted impacts of global climate change is the
desynchronization of biotic developmental patterns---such as as the inability of forests to
migrate as quickly as other aspects of their ecological niches---that have remained
coherent for millennia. This de-coupling between various elements of an ecosystem is
one of the most unpredictable and disruptive results of abrupt climate change. As
environmental scientists, Jeff Price (California State Univ., Chico) and Terry Root
(Stanford) state it, when discussing the impact of mean temperature increase
\cite{Pric06a}:
\begin{quote}
As many tree species are long-lived and migrate slowly they would be expected
to slowly colonize to the north of their range (or up in elevation), while at the
southern edge of their range their rate of reproduction slows and finally stops.
Even once a species stopped reproducing, the habitat may not undergo much
compositional change until the existing community dies out. Thus, it could take
decades to centuries for species in some vegetative communities to be replaced
by others. As increased temperatures and drought stress plants, they become
more susceptible to fires and insect breaks. These disturbances will likely play a
large role in the conversion of habitats from one type to another. There could
very well be instances where the existing plant communities are lost to
disturbance but climatic conditions and migration rates limit the speed by which
a new vegetative community replaces the original. Thus, some areas may
transitionally be replaced by grasslands, shrublands and, especially, by invasive
species. The probability of these transitional habitats may very well be increased
with abrupt climate change.
\end{quote}

Unfortunately, insects respond to changes in their thermal environment much faster than
their hosts, either through migration, adaptation, or evolution. Under the stress of abrupt
climate change the only short-term limit on their increasing populations may be their near
total elimination of suitable hosts. In short, trees only adapt slowly to changing
conditions, while insects can disperse widely and adapt much faster to abrupt
environmental changes. One conclusion is that the static, architectural view of Fig. 1
needs to be augmented to indicate the wide range of time scales involved. Table I gives
rough estimates of the times over which various feedback loop components and their
interactions adapt.

\begin{table}[tbp]
\begin{tabular}{|l||c|c|}
\hline
\multicolumn{3}{|c|}{Time Scales} \\
\hline
\hline
Component & Character & Years \\
\hline
\hline
Climate     & Season                   & $1$                 \\
            & Temperature              & $10^2$              \\
            & Glaciation               & $10^4$              \\
\hline
Forest      & Migration                & $10^2-10^3$         \\
\hline
Tree        & Infestation response     & $10^{-2}$           \\
            & Death due to infestation & $10^{-1}$           \\
            & Life cycle               & $10^2$              \\
            & Evolution                & $10^4-10^5$         \\
\hline
Bark Beetle & Tree nesting             & $10^{-2}$           \\
            & Migration                & $10^{-1} - 10^{-2}$ \\
            & Life cycle               & $1$                 \\
            & Adaptation               & $10$                \\
            & Evolution                & $10^2$              \\
\hline
\end{tabular}
\caption{Time Scales in Entomogenic Climate Change.}
\label{tab:TimeScales}
\end{table}

\section{The Tree's Perspective}

While it is clear that under extreme conditions phytophagous insects and their associated
microorganisms can quickly gain the advantage against host trees, it is also true that trees
have evolved effective defense mechanisms. For example, in their defense against bark
beetles there are two recognized components: the \emph{preformed resin system} and the \emph{induced
hypersensitivity response}. Once a beetle bores through the outer tree bark into the inner
tissues, resin ducts are severed and its flow begins. A beetle contends with the resin flow
by removing resin from its entrance hole. Trees that are sufficiently hydrated often
manage to ``pitch-out'' the invader through sufficient flow of resin. In some conifer
species with well-defined resin-duct systems, resin is stored and available for beetle
defense. The \emph{monoterpenes} within the resin also have antibiotic and repellent properties
to defend against beetle-associated fungi \cite{Nebe93a}.

The induced hypersensitivity response is usually a secondary defense system; it is also
known as \emph{wound response}. It produces secondary resinosis, cellular dessication, tissue
necrosis, and wound formation---essentially a tree's attempt to isolate and deprive
nutrition to an invading organism. In species without well-defined resin-duct systems it is
often a primary defense mechanism. In both cases these defense strategies are very
susceptible to variations in temperature and available moisture. Their efficacy also varies
with different beetle species \cite{Nebe93a}.

A series of extremely warm summers in Alaska, starting in 1987, resulted in the
dispersal of spruce bark beetles on the Kenai Peninsula when trees were water
stressed and greater beetle brood sizes survived the winter. This also halved
beetle development times from a two-year to a one-year cycle. The result was a
major increase in beetle activity that grew every year until most of the mature trees were dead \cite{Symo05a}. Since winter survivability and the number of eggs laid by bark beetles is
directly correlated to ambient temperature \cite{Lomb00a}, it is no surprise that similar increases in
yearly beetle population cycles have been observed throughout the western states and
provinces as warming and local drought conditions have persisted \cite{Wagn04a}.
As Table 1 makes clear the relative time scales for increased infestation rates, and
subsequent adaptive tree response, can put host trees at a serious disadvantage with
regard to even the short-term effects of climatic warming.

\section{Bark Beetles: Known and Unknown}

While many of the issues concerning bark beetle behavior and control can be generalized
to other insect groups, many others cannot. With this limitation in mind, we must narrow the
discussion to bark beetles to complete our analysis of the feedback loop. There are
several essential questions that dominate the study of bark beetles, many of which have
been pursued in an attempt to define viable infestation control strategies. These are:
\begin{enumerate}
\item \emph{The Pioneer Beetle}: Exactly how does a new beetle generation find new suitable
	host trees? Is the process merely random or are host-mediated attractants
	involved? Is one adaptation universal to all Scolytidae or are different processes
	used by different species?
\item \emph{Communication}: How do bark beetles communicate in order to mate, defend
	territory, coordinate tree-attack, and reduce competition within the host?
\item \emph{Beetles, Trees, and Microorganisms}: What controls the symbiotic micro-ecology
	between host trees, beetles, and the microorganisms that mediate between them?
\end{enumerate}

While humans have always been in a competitive interaction with insects, this has largely
been a stalemate. Insects have readily adapted to many of our control strategies and some
of our most effective defenses have had a tendency to backfire. In the case of the current
North American bark beetle invasions, attempts at intervention are proving mostly
negligible. The Canadian Forest Service, in response to the mountain pine beetle invasion
of British Columbia, thinned healthy forests, cut down and burned infested trees, and set
out beetle traps while hoping for a deep freeze that never came. Micheal Pelchat
(Canadian Forest Service) describes what happened: ``We lost. They built up into an army
and came across \cite{Stru06a}.'' Though ineffective, it is sobering to realize that such measures still
constitute our only defensive arsenal as the beetles move into the boreal forests.

\subsection{Chemical Micro-ecology}

Over the past thirty years many hundreds of scientific papers have been published on
bark beetles. Among reported observations, the majority focused on beetle chemical
ecology. A much smaller percentage addressed alternative aspects of beetle biology and
their relationship to the environment. In some ways this has been, for good reason, due to
the growth of the field. Chemical ecology has been one of the major successes in 20$^{\mathrm{th}}$
century entomology. As Edward O. Wilson (Harvard) states \cite{Eisn03a}:
\begin{quote}
The discipline that came into existence as a result of the pointillist studies by
Eisner, Jerrold Meinwald, and a few other pioneers was chemical ecology. Its
importance arises from the fact that the vast majority of organisms---surely more
than 99 percent of all species when plants, invertebrates, and microorganisms are
thrown in---orient, communicate, subdue prey and defend themselves by
chemical means.
\end{quote}
As this emphasizes, much of the living world communicates through chemical
signaling---intentional or inadvertent---especially through those compounds exchanged
between members of the same species, called \emph{pheromones}, or through those chemical
cues emitted by a prey source for the benefit of the predator, called \emph{kairomones}.
Chemical ecology is the study of these compounds that attempts to unravel and map this
extensive chemical
language through analysis of both chemical compounds and observation of the behavior
of living organisms correlated to them. It also seeks to discover how these compounds are
created and how to synthesize them for possible manipulation of their creators---such as
bark beetles \cite{Agos92a}.

The conventional and widely held chemical ecology model for bark beetle-tree
interactions is easily summarized. Like many other insects, bark beetles manufacture
communicative pheromones from molecular constituents that they draw from host trees. Some species of beetle are specialists that prefer a single species of tree while others have adapted to a range of different tree species.
Different species can also favor different host conditions: live trees, weak or dying trees, or recently fallen timber. The breeding site within a host also varies, with different species taking up residence in either the lower or
upper trunk, with still others preferring the crown. Presumably, this localization evolved
to reduce competition between species, allowing a diversity to co-exist in a single tree.
Many of the vast number of different Scolytidae species have evolved to maintain a non-lethal relationship to their hosts. Others have evolved to normally colonize dead or dying trees at normal low-population densities but can colonize living trees when populations reach high levels.
Finally, there are the primary bark beetle species that normally kill their hosts. These species, that are usually the most destructive, can stage a mass attack and use aggregation pheromones between beetles to trigger this behavior. 

\subsection{Pioneer Beetle: Infestation linchpin}

An attack begins with the pioneer beetle that locates, by means not yet elucidated, and lands
upon a suitable host. Others join this beetle, all soon boring through the outer bark into
the phloem and cambium layers where eggs are laid after mating. Within the resulting
galleries that house the adult beetles and their eggs, the larvae hatch, pupate, and undergo
metamorphosis into adulthood. In this way, they spend the largest fraction of their life-cycle (anywhere between 2 months to two years depending upon species and geographic location) inside a tree. This new generation emerges from the bark and flies
away to seek new host trees. The widely held hypothesis is that the pioneer also attract
other beetles to the host through a pheromone signal. In some species the pioneer is male
and, in others, female. Each new beetle that is attracted to the host subsequently
contributes to the general release of the aggregation pheromone. It is also theorized that
the aggregation pheromone has an upper limit beyond which attracted beetles will land
on adjacent trees rather than the initial host, since high concentrations would indicate
over-use of the available host resources. Resident bacteria within the beetles may facilitate
the production of aggregation pheromones. The aggregation pheromone of one species also
tends to repel other species \cite{Agos92a}.

As new research has filled in the gaps of the chemical ecology model, it has become
clearer that it is often over-simplified. There are a large number of nonchemical
mediating factors that are both independent of pheromone signaling and directly affective
of its role. The model tends to emphasize the chemical and olfactory mechanisms of bark
beetles and downplays, or simply ignores, a large array of other factors. In some ways,
regarding its dominance in the study of bark beetles, the chemical paradigm has become a
victim of its own success. While proving to be a seemingly inexhaustible well of new
hypotheses about the chemical intricacies of these creatures and their relationship to host
trees, it has failed to answer many of the central questions about their reproductive and
mating behavior and so their infestation dynamics.

One hope has been that understanding bark beetle chemical ecology would lead to its
manipulation and eventually to a viable forestry management tool. Much to our loss,
nothing of the sort has been forthcoming. This largely derives from the sheer complexity
of the insect-tree micro-ecology and how far away we are from a sufficient understanding
of mechanisms and interactions. The two major contributions of chemical ecology
research to control measures have been those of pesticides and pheromone trapping. Most
biologists appreciate that pesticides have a very limited role in controlling insect
infestations at the scales in question. Pheromone traps are one of the essential tools of
field research in entomology, but adapting them for large-scale control has been
controversial at best; see Borden 1997 \cite{Bord97a} for an overview. Some researchers and
chemical manufacturers assert that they have a positive effect in collecting, condensing,
and re-directing of beetle populations \cite{Schl01a}. Other researchers, however, claim that these
effects are inconclusive and, worse, in some cases may exacerbate negative conditions
\cite{Wesl92a}. In any case, effective traps and synthetic pheromone production are costly and their
toxicity is undetermined.

These issues are illustrated by the only large-scale study of the effectiveness of
pheromone trapping. In 1979 a massive outbreak of the European spruce bark beetle (\emph{Ips
typographus}) spread through the forests of Norway and Sweden. A large trap-out
program was implemented in an attempt to counter the invasion of the Scandinavian
spruce forests. 600,000 baited pheromone traps were placed throughout the infested
forest. It yielded a ``capture'' of three billion beetles in 1979 and four billion in 1980.
Despite what appears to be substantial numbers of captured beetles, ultimately the
infestation was devastating. It appeared to have simply run its course. Trap effectiveness
could not easily be evaluated. No consensus conclusions were drawn as to whether the
deforestation would have been worse or better without them. Notably, despite increased
infestation, no pheromone intervention of this scale has been attempted since.
Unfortunately, there is still no clear picture of what was accomplished \cite{Agos92a}.

Given the differing scientific opinions and the as-yet undemonstrated benefits,
pheromone trapping, the control strategy that derives from the chemical ecology
paradigm, remains controversial. The apparent ineffectiveness of large-scale pheromone
trapping, though, illustrates one of the central unanswered questions of the pioneer beetle.
An underlying assumption of chemical ecology is that pheromones are the primary
attractant for beetles seeking new hosts, but this remains a hypothesis. While many
researchers believe that attraction is olfactory, others propose that visual cues are key for
some species \cite{Camp06a}. Importantly, forestry management policy is based largely on the
chemical ecology hypothesis that olfaction is dominant. It has never been definitively
proven, however, and, for a number of reasons, it is unlikely to be. Stated simply, foraging
insects most likely use whatever cues are the most accurate and easily assessed under varying
circumstances. To assume otherwise is to go against the common logic that living
systems evolve multiple survival strategies to cope with environmental complexity.

\subsection{More Pieces of the Puzzle}

Other aspects of bark beetle behavior also remain mysterious. One of the most curious
observations concerns the role lightning-damaged trees play in sustaining populations of
some bark beetle species. For example, lightning plays an essential role in the ecology of
southern pine beetle (\emph{Dendroctonus frontalis}) populations in the Gulf Coastal Plain.
Notably, infestations are observed to begin in the spring when beetle populations are usually
low and lightening is frequent. Some years, records show that 75 percent of beetle
infestations were associated with lightning strikes. In addition, the number of beetles in a
struck tree averages nearly four times that of an unstruck, but infested tree. While there
are obvious and dramatic chemical changes that occur after such a strike, none explain
this extraordinary behavior, especially how beetles of several different species can be
attracted from a substantial distance to a single struck tree. It appears that no substantial
hypothesis has been put forward \cite{Hodg71a}. This is one aspect of bark beetle
behavior that begs for a novel perspective and, therefore, a new explanatory mechanism.

Another more recent curious observation concerns how some bark beetle infestations are
associated with certain rock and tree stand types. A recent study, using landscape-level
geographic analysis of the Panhandle National Forests of northern Idaho and eastern
Washington, shows a correlation of Douglas-Fir beetle (\emph{Dendroctonus pseudotsugae})
infestation rates with certain geologic strata and forest stand types. While the authors of
this study speculate that perhaps the correlation is due to variations in nutrient stress that
impact tree vulnerability, to date no research has verified this \cite{Garr03a}. In fact, the hypothesis that this effect can be correlated to nutrient stress says nothing about the nature of the tree vulnerability cues that might be communicated to beetles.

One of the most exciting areas of bark beetle research analyzes the complex
micro-ecological dynamics between beetles, various forms of fungi, and diverse species of
mites. For example, we now know that almost every state and federal information
website gives an incomplete description of how bark beetles kill conifers. The story is
that it is a fungus, carried by the beetle, which infects a tree's vascular system, choking
off the flow of nutrients. While it is true that a compromised vascular system can
ultimately kill a tree, the websites describe the relationship between blue-stain fungus
(genus \emph{Ophiostoma}) and the beetles as if there is only one organism and only one such
species of fungus involved. The truth is that there is no clear consensus about what
actually kills host trees. The emerging picture is a very complex one that involves many participatory agents. In fact, many different species of
\emph{Ophiostoma} and many other genera of fungi---including \emph{Entomocorticium},
\emph{Ceratocystiopsis}, \emph{Trichosporium}, \emph{Leptographium}, and \emph{Ceratocystis}---are involved \cite{Pain97a}.
There are probably other interspecific interactions between different species of beetles that are
also significant \cite{Byer81a}. Moreover, mites that live on the beetles also
carry the fungi \cite{Hofs06a}.

The resulting picture is of a constantly shifting dynamic that includes fungi, mites,
beetles, and trees. Moreover, the relationships involve different modes of symbiosis:
parasitism, mutualism, and commensalism. Each of these, in turn, affects infestation
dynamics in different ways. This might seem like a hopelessly complicated
micro-ecology. However, it is this very diversity that leads to the hope that
unraveling its complexity may contribute towards new biological control regimes
\cite{Klep01a}.

Since these mysteries and micro-ecological dynamics cannot be adequately explained by the familiar aspects of the chemical ecology paradigm, there is a need to look for other explanatory mechanisms that might bridge the gaps in our understanding or offer novel insights.

\subsection{The Bioacoustic Ecology Hypothesis}

The motivation to search out novel control regimes is clearly a response to the serious
limitations that chemical control strategies have faced. This has necessitated both a
search for entirely new areas of investigation, such as the previously mentioned
micro-ecological dynamics, and a resurgence in areas of research that have received minimal
attention in the past.

One of the more neglected research domains regarding bark beetles concerns their
remarkable bioacoustic abilities. The sound producing mechanism in many bark beetles is
most likely a \emph{pars stridens} organ that functions as a friction-based grating surface.
In \emph{Ips confusus} beetles it is located on the back of the head and stroked by
a \emph{plectrum} on the under side of the dorsal anterior edge of the prothorax.
In other species (\emph{Dendroctonus}) the pars stridens is located on the surface
under the elytra and near the apices and sutural margins. Another is found in some
species on the underside of the head. All three of these sound generating organs
produce a variety of chirps that range from simple single-impulse clicks to a range
of different multi-impulse chirps. These also differ between genders of the same
species and between different species probably due to subtle differences in the
sound producing mechanisms. Collectively, all of the sounds and their associated
mechanisms are referred to as \emph{stridulation}, the most common form of sound
production made by various forms of beetle \cite{Barr69a}.

In monogamous species of bark beetles, such as those of the genus
\emph{Dendroctonus}, the female is the pioneer and does not possess the complex
pars stridens that the male uses. The opposite is true of the polygamous \emph{Ips}
genus where the male is the pioneer and the female the one with a pars stridens
organ. In both cases, however, the pioneer is also known to produce simpler forms
of communicative signaling using other, less understood sound generating
mechanisms. We do know that there are at least 14 different types of stridulatory
organs amongst adult beetles in thirty different families \cite{Wess06a}. So the potential
for undiscovered varieties of such mechanisms among Scolytidae is high.

Past research suggested that sound making and
perception in bark beetles was of low significance compared to their chemical-signaling
mechanisms. In fact, of the studies that dealt with their acoustic behavior, over half
concentrated on the relationship of sound generation to chemical signaling. These include
the role stridulation sound-making has in controlling attack spacing between entry points
in the host \cite{Byer89a} or the triggering of pheromone release between genders
\cite{Rudi76a}. The resulting view is that bark beetles use a combination of chemical
and acoustic signals to regulate aggression, attack on host trees, courtship, mating
behavior, and population density.

While the dual behavioral mechanisms of scent and sound are largely
inseparable, it is usually assumed that bark beetles use chemical messages for
communication at a distance while reserving acoustic signals for close-range
communication. However, this distinction remains hypothetical. We do not have a
definitive understanding of how far either their pheromones or sound signals can travel,
let alone a full appreciation of the diverse forms of acoustic signaling that they may
employ. We do know that both communication mechanisms are used after beetles have
aggregated on a host and that one form of signaling can evoke the other.

Bark beetles exhibit an amazing array of complex social behaviors---such as, group
living, coordination of mass attack, the necessity for mass infestation to effectively
counter host defenses, signaling to reduce intraspecific competition, and the collective
occupation of nuptial chambers by polygamous species. This complexity implies that the
coupling between communication mechanisms is significant. Despite the importance that
their acoustic communication must have for overall survival and environmental fitness,
there are still no published studies on their sound reception mechanisms or any
identification of hearing organs.

The broad neglect of bark-beetle bioacoustic behavior has also led to a lack of follow-up
on the proposal that host trees themselves produce acoustic cues that also attract pioneer
beetles. Perhaps the earliest proposal dates back to 1987, when William Mattson
and Robert Haack
(USDA, Forest Service) speculated that cavitation events in trees might produce
acoustic signals audible to plant-eating insects \cite{Matt87a}. Cavitation occurs in trees by breaking
of the water columns in the conducting xylem tissue of leaves, stems, and trunks. The
assumption has been that the sounds are vibrations coming from individual cells
collapsing, which is due to gradual dehydration and prolonged water stress. While
cavitation produces some acoustic emissions in the audible range, most occur in the
ultrasound range. In fact, counting ultrasonic acoustic emissions from cavitating xylem
tissues is a widely accepted monitoring practice used by botanists to measure drought
stress in trees. Despite its common usage in botany, there has been very little study as to
the actual generating mechanism. For the most part, it is merely a statistical measuring
tool and the correlation between the incidence of cavitations and drought stress, an
accepted fact \cite{Jack96a}.

This proposal requires, of course, that bark beetles be able to perceive the
drought-stressed tree's ultrasonic acoustic emissions. At the present time, while insect
sound-making mechanisms are fairly well understood, this is not the case with their auditory
organs. Nonetheless, every year the list of insect species shown to have ultrasonic hearing
grows. It now includes many species of butterflies and moths (\emph{Lepidoptera}), mantids
(\emph{Mantodea}), grasshoppers and crickets (\emph{Orthoptera}), flies (\emph{Diptera}), and net-veined
insects (\emph{Neuroptera}).

There has also been increasing investigation of interspecific sensing
in the ultrasonic range, such as the influence of bat echolocation on the evolution of
moths and butterflies. In fact, it appears that much of the evolution of ultrasonic hearing
in flying insects has been driven by this essential predator-prey relationship. These
insects have evolved a startle response to the presence of echolocating bat chirps and take
avoidance measures by suddenly dropping in flight \cite{Hoy89a}.

Interestingly, despite their being the largest insect order, there have been only two kinds
of beetles (Coleoptera) discovered to have tympanum-hearing organs: scarabs
(Scarabaeidae) and tiger beetles (Cicindelidae). This appears to be more a matter of lack
of study than a general characteristic of the order. Researchers believe that many other ultrasound sensitive beetles will soon be discovered \cite{Forr97a}.

\subsection{Testing the Hypothesis}

Recent fieldwork by one of us (DDD) focused on sound production by the pinion engraver
beetle (\emph{Ips confuses}). Sounds were recorded within the interior phloem layer of
the trees, often adjacent to beetle nuptial chambers. A rich and varied acoustic ecology
was documented---an ecology that goes beyond the previously held assumptions about
the role of sound within this species \cite{Dunn06a}. Another important observation was
that much of the sound production by this species has a very strong ultrasonic component.
Since communication systems seldom evolve through investing substantial resources into a
portions of the frequency spectrum that an organism cannot both generate and perceive
\cite{Dunn06b}, this raised the question of whether or not bark beetles have a complementary
ultrasonic auditory capability.

Recent laboratory investigations by Jayne Yack (Neuroethology Lab, Carleton University)
have also revealed ultrasound components in some bark beetle signals and indirect
evidence that the beetles might possess sensory organs for hearing airborne sounds. One
possible implication that arises from the combination of these laboratory and field
observations is that various bark beetle species may possess organs capable of hearing
ultrasound for conspecific communication and are therefore preadpated for listening to
diverse auditory cues from trees \cite{Yack06a}.

If further evidence of ultrasonic perception can be verified in this and other bark beetle
species, then a number of interesting possibilities arise. It has been a working assumption
among entomologists studying Scolytidae that bark beetles are not under predation
pressure from insectivorous bats. The claim is that bark beetles do not fly at night.
This would mean that the most likely explanation for bark beetles evolving an ultrasonic
hearing capability is not applicable since it would be, in familiar evolutionary terms,
an unnecessary adaptation. Thus, it would appear that such an adaptation must have evolved
to monitor environmental sound cues, such as cavitation acoustic emissions, or a previously
unknown intraspecific signaling system in the ultrasonic range. If verified, this would
contribute substantially to an improved understanding of the role that sonic communication
plays in the development, organization, and behavior of bark beetles---a key and previously
unsuspected role.

\subsection{Multimodal Sensing, Communication, and Social Organization}

In the overall scheme of entomogenic climate change, the complex feedback loop appears
to turn critically on the bioacoustic and chemical-mediated interactions between beetles
and trees. Given this, where else might we look for control methods?

While receiving little or no concerted attention, there is one area of possible bark beetle
research that warrants discussion since it could have important impacts on the design of
new control methods. While bark beetles appear to have a complex communication
system that uses both chemical and acoustic forms of signaling, the question of how
complex their social organization might be has seldom been asked. Again, behaviors such
as group living, coordination of mass attack, the necessity for mass infestation to
effectively counter host defenses, signaling to reduce intraspecific competition, and the
collective occupation of nuptial chambers by polygamous species, seem to imply that
some level of rudimentary social awareness is implicated in bark beetle behavior and
necessary for survival of the various species. How far this resembles more familiar forms
of insect eusocial behavior remains an open question.

In recent years the important role of insect communication through vibrational substrates
has become clear. Despite the dominance of chemical ecology, one now reads that of insect
species using mechanical communication (sound in air, ripples on water, and so on)
\cite{Cocr05a}: ``92 percent use substrate vibrations alone or in concert with other
forms of mechanical signaling.''
Reginald Cocroft (University of Missouri-Columbia) has hypothesized that for many
group-living insects that feed on plants, substrate vibrational signaling is an essential
aspect of how they exploit environmental resources. He suggests that there are at least
three different kinds of challenges to these insects that are met by communication
through plant substrates: locating and remaining in a conspecific group, locating food,
and avoiding predators \cite{Cocr01b}. 

We are most familiar with the complex eusocial systems and
related (and essential) communication systems of bees, wasps, ants, and termites. The
above acoustic fieldwork has led us to conclude that there must be a larger range of forms
of insect sociality and therefore means of organizational communication. More precise
understanding of these forms of social organization may improve our ability to design
better control systems, whether these are chemical, acoustic, or biological.

In investigating the sound communication of pinion engraver beetles, one conclusion
became inescapable. The phloem and cambium layers of pinion trees are an amazingly
effective medium for acoustic communication. Individual stridulations can carry for
several feet within the tree bark interior, most of which are inaudible to an outside
(human) listener or to sensitive recording apparatus. Moreover, the diverse sounds made
by the pinion engraver beetles appear to effectively match the combination of cellulose,
fluids, and air that comprise these tissue layers. In communication theory terms, there is
an effective impedance match between sound generator and the transmission medium.
These layers are also an appropriate acoustic medium for several other invertebrate sound
makers. The field recordings reveal that the tree interior is a rich and teaming world of
sound \cite{Dunn06b}---its own bioacoustic ecology.

These observations raise an important issue not addressed by previous bark beetle
bioacoustic research. A very diverse range of sound signaling persists well after
the putatively associated behaviors---host selection, coordination of attack, courtship,
territorial competition, and nuptial chamber excavations---have all taken place. In
fully colonized trees the stridulations, chirps, and clicks can go on continuously for
days and weeks, long after most of these other behaviors will have apparently run their
course. These observations suggest that these insects have a more sophisticated social
organization than previously suspected---one that requires ongoing communication
through sound and substrate vibration.

The results in both bioacoustics and chemical ecology strongly suggest bark beetle
communication is largely multimodal and that both pheromone and mechanical signaling
are interwoven. A growing appreciation in many fields of biology has emerged that
animal signals often consist of multiple parts within or across sensory modalities. Insects
are not only an example of this observation, but they possess some of the most surprising
examples of multicomponent and multimodal communication systems \cite{Skal05a}. Sometimes these
different components or modalities signal different information and sometimes they are
redundant. For example, it was assumed that bee communication, through either the
famous ``waggle dance'' or associated sounds from wing vibrations, communicated
different informational content during display. More recently, experiments with robot
bees determined that these systems are largely redundant, most likely a strategy for
reducing transmission errors \cite{Mich99a}. Collectively, these observations reinforce the opinion
voiced many years ago by entomologist Philip Callahan (University of Florida) \cite{Call81a} that:
``as long as sound is studied in one corner of the lab and scent in another, the mechanisms
of these sound-modulated scent molecules will not be understood ...''.

\section{Conclusion: Closing the Loop}

The eventual impact that insect-driven deforestation and global climate change will have
on the Earth's remaining forests ultimately depends on the rate at which temperatures
increase. The impacts will be minimized if that rate is gradual, but increasingly disruptive
if the change is abrupt. Unfortunately, most climate projections now show that a rapid
temperature increase is more likely \cite{Wats01a}. The current signs of increasing insect
populations at this early stage of warming does not portend well for forest health in the
near future. The concern is exacerbated, since we have limited countermeasures under
development.

One conclusion appears certain. Extensive deforestation by insects will convert the
essential carbon pool provided by the Earth's forests into atmospheric carbon dioxide.
Concomitantly, the generation of atmospheric oxygen by trees will decrease. Most
immediately, though, as millions of trees die, they not only cease to participate in the
global carbon cycle, but become potential fuel for more frequent and increasingly
large-scale fire outbreaks. These fires will release further carbon dioxide into the
atmosphere and do so more rapidly than the natural cycle of biomass decay. The interaction
between these various components and the net effect is complicated at best---a theme that
runs through links in the entire feedback loop.

An example of this is how boreal forest fires affect climate \cite{Rand06a}. A
constellation of substantially changed components (lost forest, sudden release of gases,
and the like) leads, it is claimed, to no net climate impact. The repeated lesson of
complex, nonlinear dynamical systems, though, is that the apparent stability of any part
can be destabilized by its place in a larger system. Thus, one needs to evaluate the lack of
boreal fire-climate effects in the context of the entire feedback loop.

Taken alone, the potential loss of forests is of substantial concern to humans. When
viewing this system as a feedback loop, though, the concern is that the individual
components will become part of an accelerating positive feedback loop of sudden
climatic change. Such entomogenic change, given the adaptive population dynamics of a
key player (insects), may happen on a very short time scale. This necessitates a shift in
the current characterization of increasing insect populations as merely symptomatic of
global climate change to a concern for insects as a significant generative agent.

While current research programs will continue to contribute important insights on
chemical communication and associated behavior of plant-eating insects, hard-won
experience suggests it is increasingly less plausible that the chemical ecology paradigm
alone will be the source for effective intervention strategies, as originally hoped. We
believe that alternative approaches will contribute fresh insights and suggest innovative
mechanisms for detection, monitoring, and control. Most importantly, we conclude from
the complexity of the constituents and interactions in the feedback loop that there must be
greater support for interdisciplinary approaches. At a minimum, the problem we
described requires a more comprehensive understanding of insect multimodal and
multicomponent communication and its rich ecological context. These then must be
evaluated in the larger frame of entomogenic climate change.

In addition to concerted research in bioacoustics, micro-ecological symbiosis and
dynamics, and insect social organizations, these areas, in conjunction with the field of
chemical ecology, must be integrated into a broader view of multiscale population,
evolutionary, and climate dynamics. In this sense, the birth of chemical ecology serves as
an inspiration. It grew out of an interdisciplinary collaboration between biology and
chemistry. It is precisely this kind of intentional cooperation between disciplines---but
now over a greater range of scales---that will most likely lead to new strategies for
monitoring and defense against what seems to be a growing threat to the world's forests
and ultimately to humanity itself.

\section*{Acknowledgments}

The authors thank Dawn Sumner, Jim Tolisano, Richard Hofstetter, Jayne Yack,
Reagan McGuire, and Bob Harrill for helpful discussions. This work was partially
supported by the Art and Science Laboratory via a grant from the Delle Foundation
and the Network Dynamics Program, funded by Intel Corporation, at UCD and the
Santa Fe Institute.

\bibliography{ref}

\end{document}